\documentclass[a4paper]{jetpl}
\usepackage[russian]{babel}
\rus
\title{
Renormalization in Lorentz-Abraham-Dirac Equation, Describing Radiation Force in Classical Electrodynamics (in Russian). 
Перенормировка уравнения Лоренца-Абрагама-Дирака для радиационной силы в 
классической электродинамике}
\rtitle{Renormalization in the Lorentz-Abraham-Dirac equation} 
\author{Igor V.\,Sokolov\/\thanks{igorsok@umich.edu}}
\address{University of Michigan, 2455 Hayward Str., MI48109 USA}
\abstract{
While he derived the equation for the radiation force, Dirac (1938) mentioned a possibility to use different choices for 
4-momentum of an emitting electron. Particularly, the 4-momentum could be non-colinear to the electron 4-velocity.
This ambiguity in the electron 4-momentum allows us to assume that the mass of emitting electron may be an operator, 
or, at least, a 4-tensor instead of being the usually assumed scalar, 
which relates the 4-velocity of a bare charge to the total momentum of a dressed point electron, 
the latter being a total of the momentum of the bare electron and that of the own electromagnetic field.
 
On applying the re-normalization procedure to the mass operator, we arrive at an interesting dichotomy. The first choice 
(more close to traditional one) ensures the radiation force to be orthogonal to 4-velocity. In this way the re-normalization results in 
the Lorentz-Abraham-Dirac equation or in the Eliezer equation. However, the 4-momentum of electron in this case is not well defined: 
the equality in the relativistic entity 
$({\cal E}/c)^2=m^2c^2+p^2$ appears to be broken and even the energy is not definite positive. 
The ill-defined energy is an underlying reason for the run-away solution, in which the electron in the absence of external field loses 
the radiation energy and acquires the kinetic energy in the same time.
 
The other choice is to require the radiation force to be orthogonal to the 4-momentum (not to the 4-velocity). In this case the energy and 
momentum are well-defined and obey the relationship $({\cal E}/c)^2=m^2c^2+p^2$. Remarkably, the equations of a particle's motion in this case 
differ significantly from all the known versions. They appear to be well founded. They are simple, easy to solve, and can be applied to 
simulate the particle motion in the focus of an ultra-bright laser.  
}

\begin{document}
\maketitle
В работе Дирака о радиационной силе в классической электродинамике предусматривалась возможность введения 4-импульса излучающего электрона, 
неколинеарного с 4-вектором его скорости. Тем самым фактически предлагалась перенормировка массы как оператора, соотносящего указанные векторы. Там же отмечалась 
неединственность выбора, приводящего к уравнению Лоренца-Абрагама-Дирака (ЛАД).  Как показано  в настоящей работе,  перенормировка уравнения ЛАД при дополнительном 
требовании, чтобы перенормированные энергия-импульс подчинялись такому же релятивистскому соотношению, как и обычные, (${\cal E}^2={\bf p}^2c^2+m^2c^4$),  приводит 
к существенной модификации уравнения ЛАД. Свойства нового уравнения: оно строже, чем уравнение ЛАД, будучи свободно от пороков последнего; оно проще, чем известное 
приближение для уравнения ЛАД; оно представляется предпочтительным для численного моделирования процессов в полях сверхсильных лазерных импульсов.

Движение электрона в фокусе мощного лазера, такого как HERCULES (University of Michigan)  ~\cite{bib:V}, в основном контролируется излучательными потерями энергии и 
импульса ~\cite{bib:B}. Проблема возможно более точного описания такого движения в рамках классической электродинамики и возможно более эффективного численного 
интегрирования уравнений движения для систем многих частиц (плазмы) в значительной степени остается открытой и актуальной. 

Вообще говоря, при этом необходимо решать уравнение Лоренца-Абрагама-Дирака (ЛАД).  Обзор теории уравнения ЛАД  дан в \cite{bib:K}, достаточно полный полный список 
современных публикаций имеется в ~\cite{bib:R}, и здесь отдельно цитируются только те работы, которые кажутся уместными в обсуждаемом контексте.  Прямое численное 
моделирование для уравнения ЛАД, у которого ``почти все'' решения улетают на бесконечность (runaway) за малое время,  осуществлено в ~\cite{bib:H}. При этом примененный 
метод (интегрирование вспять по времени), хотя изящен, но вряд ли пригоден для многочастичных систем, что лишний раз подчеркивает незаконченность  теории, несмотря на 
значительные усилия в этом направлении.  Известны приближенные версии уравнения ЛАД, несомненно, лучшая из них описана в ~\cite{bib:L} (см уравнение (76.3)). Все 
приближенные версии не свободны от недостатков (см обсуждение и ссылки в ~\cite{bib:K}), таких, например, как неточное сохранение энергии-импульса в системе внешнее 
поле-частица-излучение. 

В настояшей работе используется идея, намеченная Дираком  ~\cite{bib:D}, но  не разработанная.  Именно, предлагается перенормировка массы электрона, но не как скаляра, 
соотносящего полную энергию одетого электрона с энергией покоя голого заряда и его электростатической энергией - а как оператора, соотносящего 4-импульс одетой частицы 
с 4-скоростью голого заряда, то есть с током. При этом как равноправная альтернатива  уравнению ЛАД, а не как приближение, возникает система уравнений, которая кажется 
более пригодной, уж по крайней мере для численного моделирования. 

Подчеркнем, что здесь обсуждаются только точные законы сохранения и удовлетворяющие им уравнения движения, но к самим уравнениям применение понятия ``точные'' вряд ли 
оправдано. Сама концепция радиационной силы в ~\cite{bib:L} (как и уравнение ЛАД) вводится как приближенная. Вне пределов приближения излучение высокоэнергичного фотона 
приводило бы к значительному изменению импульса электрона $\Delta p^i$, которое не может быть сведено к импульсу силы $(f^i)_{\rm rad}\Delta t$ за сколь угодно малый 
интервал времени $\Delta t$.  Столь далеко выходящий за пределы классической электродиамики случай здесь не рассматривается.  

\textbf{В пренебрежении излучением} движение электрона во внешнем электромагнитном поле описывается следующими уравнениями ~\cite{bib:L}:
\begin{equation}
\dot{p}^i=\frac{e}{c}F^{ij}\dot{x}_j,\qquad
\dot{x}^i=m^{-1}{p}^i,\label{eq:1}
\end{equation}
где $F^{ij}$ - тензор поля, $p^i$ и $x^i$ - 4-векторы механического импульса и координат электрона, $e=-|e|$ и $m$ - его заряд и масса покоя, $c$ - скорость света. Точка 
обозначает дифференцирование по собственному времени, $\tau$. 4-векторы и 4-тензоры перемножаются по обычным правилам с метрическим тензором, 
$g^{ij}={\rm diag}(1,-1,-1,-1)$ . Динамика поля дается уравнением для тензора энергии-импульса, $T^{ij}$:
\begin{equation}
\frac{d}{d\tau}\int{{T}^{i0}dV}=-\frac{e}{c}F^{ij}\dot{x}_j,\label{eq:2}
\end{equation}
так что суммарные энергия и импульс сохраняются: ${p}^i+\int{{T}^{i0}dV}={\rm const}$. 

Система (\ref{eq:1}) - Гамильтонова,  с каноническим (обобщенным) импульсом, $P^i=p^i+eA^i/c$, что следует из представления тензора поля через 4-вектор потенциала, 
$A^i$: $F^{ij}=(\partial A^j/\partial x_i)-(\partial A^i/\partial x_j)$. Если вектор-потенциал не зависит от (циклической) координаты вдоль некоторого направления, 
$({\bf n}\cdot\nabla)A^i$, то проекция обобщенного потенциала на это направление сохраняется:  $({\bf n}\cdot\dot{\bf P})=0$. Правая часть уравнения (\ref{eq:1}) 
ортогональна как 4-импульсу, так и (колинеарной с ним) 4-скорости, что обеспечивает выполнение тождеств:
\begin{align}
\dot{x}^2&=c^2,\label{eq:3}\\
p^2&=m^2c^2, \label{eq:4}
\end{align}
здесь и далее квадрат любого 4-х вектора, $a^i$, определен согласно $a^2=a^ia_i$.

\textbf{Введение перенормировки} в уравнения (\ref{eq:1})  не является необходимостью, однако она без труда может быть проведена ~\cite{bib:L}. Сначала рассмотрим 
покоящийся электрон и окружим его точечный заряд сферической поверхностью очень малого радиуса $\epsilon$. Подсчитаем энергию электростатического поля, созданного 
зарядом электрона снаружи от указанной поверхности:  ${\cal E}_{\rm em}=\frac{e^2}{2}\int_\epsilon^\infty{\frac{dr}{r^2}}$, где $r$ расстояние до заряда. Из соображений 
релятивистской инвариатности следует выражение для 4-импульса собственного электромагнитного поля электрона, равномерно движущегося со скоростью $\dot{x}$.  Снаружи от 
окружающей заряд и движущейся вместе с ним поверхности интеграл энергии-импульса поля равен: 
$p^i_{\rm em}=\frac{e^2\dot{x}^i}{2c^2}\int_\epsilon^\infty{\frac{dr}{r^2}}$ (более длинное доказательство дано в \cite{bib:J}, Гл. 17.5). 

В пренебрежении ускорением электрона  можно суммировать импульс увлекаемого поля с импульсом заключенной внутри поверхности ``голой'' частицы: $p^i_n=m_{n}\dot{x}$. При 
этом можно ввести массу ``одетой'' частицы как  скаляр $m=m_n+e^2/(2\epsilon c^2)$, пользуясь колинеарностью  всех импульсов: 
$p^i_{em}\propto\dot{x}^i$ и $p^i_n \propto\dot{x}^i$. 

\textbf{При учете ускорения электрона} cитуация качественно изменяется. Во-первых, необходимо учитывать эффекты запаздывания в формировании электромагнитной энергии 
электрона:  для грубых оценок полагаем, что  часть энергии-импульса поля при 
$r>c\Delta \tau$ относится к значению скорости в момент времени $\tau-\Delta\tau$, тогда как скорость в текущий момент времени отнесена к интервалу интегрирования 
$\epsilon<r<c\Delta t$, и получаем: 
$$p^i_{\rm em}=\frac{e^2 }{2c^2} \left[\dot{x}^i (\tau-\Delta\tau)\int_{c\Delta\tau}^\infty{\frac{dr}{r^2}}+
\dot{x}^i \int_\epsilon^{c\Delta\tau}{\frac{dr}{r^2}}  \right],$$
или $ p^i_{\rm em}=\frac{e^2 }{2c^2} [\frac{\dot{x}^i}\epsilon-\frac{\ddot{x}^i}c ]$, если считать что  $\Delta \tau\rightarrow 0$, но при этом $c\Delta\tau>\epsilon$.  

Суммируя с импульсом голого заряда, мы находим, что с учетом эффекта запаздывания полный импульс ускоренно  движущегося электрона  не вполне колинеарен с направлением 
4-скорости голого заряда: 
\begin{equation}
p^i=p^i_{\rm em}+m_nc\dot{x}^i= m\dot{x}^i-m(\dot{x})^i_{\rm rad},\label{eq:4a}
\end{equation} 
где $(\dot{x})^i_{\rm rad}\sim\tau_0\ddot{x}^i $, $\tau_0=2e^2/(3mc^3)\approx 6.2\cdot10^{-24}\,{\rm s}$.  Связанный с ускорением дополнительный член в импульсе здесь 
и ниже обозначен как $-m(\dot{x})^i_{\rm rad}$,  поскольку для дальнейшего существенно, что в системе отсчета, в которой импульс электрона равен нулю, ток ускоренно 
движущегося (=излучающего) электрона отличен от нуля. Дополнительный член приблизительно 4-ортогонален  4-скорости, что и свидетельствует о неколинеарности импульса и 
скорости голого заряда или, что то же самое, импульса и тока. Выбор численного коэффициента в определении $\tau_0$ обеспечивает точное согласие с  расчетом Дирака 
\cite{bib:D}, который вычислял интеграл энергии-импульса для связанного электромагнитного поля электрона (о тензоре энергии-импульса для свободного поля см ниже). 
Точнее, вычислялось изменение этого объемного интеграла во времени, еще точнее - равный ему с обратным знаком поток тензора энергии-импульса через окружающую  
электрон гиперповерхность.

Однако из тех же соображений о запаздывании следует, что как бы  ни были 
точны расчеты, их результат не может дать точное значение энергии-импульса связанного электромагнитного поля электрона в момент времени $\tau$. Во-первых, вовлеченное в расчет собственное поле испущено (emanated) электроном не позднее момента времени $\tau-O(\epsilon/c)$, так что в лучшем случае вычисленный импульс относится к некоторому предшествующему моменту времени (либо размазан по некоторому интервалу времени). Во-вторых, экстраполяция сингулярного члена $\frac{e^2}{2c^2\epsilon}\dot{x}(\tau-\frac\epsilon{c})$ в  $ p^i_{\rm em}$ к текущему моменту времени дала бы погрешность порядка $m\tau_0\ddot{x}$, которая не стремится к нулю при стремлении $\epsilon$ к нулю.  

Тесно связано с отмеченным эффектом и то обстоятельство, что незначительное переопределение охватывающей электрон гиперповерхности, по которой производится интегрирование  потока энергии-импульса, приводит к кажущемуся исчезновению пропорционального  4-ускорению члена из интеграла энергии-импульса (см \cite{bib:T} или полезный обзор \cite{bib:P}). При этом все равно приходится вводить такой член в определение импульса, но уже через постулированное выражение определенного вида для импульса ``того, что находится внутри'' указанной гиперповерхности, причем произвольный коэффициент в этом выражении должен определяться из независимых соображений.  Подход весьма близок к использованному в настоящей работе.

Поскольку ускоренно движущейся электрон в классической электродинамике обязательно излучает, необходимо проследить, чтобы вычеркивание связанного поля из тензора энергии-импульса не затрагивало бы ту часть тензора, которая описывает излучение.  Такое возможно, поскольку в тензоре энергии-импульса собственного поля электрона может быть выделена относящаяся к излучению часть, 
\begin{equation}\label{eq:4aa}
T^{ij}_{\rm rad}=\frac{e^2}{4\pi r^2c^4}\kappa^i\kappa^j\left[ - (\ddot{x}(\tau))^2-(\ddot{x}^l(\tau)\kappa_l)^2\right],
\end{equation} 
\begin{equation}\label{eq:4ab}
r=[r^i-x^i(\tau)]\dot{x}_i(\tau)/c,\qquad \kappa^i=[r^i-x^i(\tau)]/r,
\end{equation}
которая сама по себе подчииняется закону сохранения во всем пространстве, за исключением мировой линии голого заряда (см выражение для $ T^{ij}_{\rm rad}$ и доказательства нижеследующих утверждений в \cite{bib:T}-\cite{bib:P}): 
\begin{equation}
\frac{\partial T^{ij}_{\rm rad}}{\partial r^j}=c\int{(\dot{p}^i)_{\rm rad}\delta^4(r^i-x^i(\tau))d\tau}.\label{eq:4b}
\end{equation}
Здесь $r^i$ - координатный 4-вектор в некоторой системе отсчета,  $\tau$ в  уравнениях (\ref{eq:4aa}-\ref{eq:4ab}) определяется для данной точки $r^i$  из условия запаздывания: $k^ik_i=0$;  $(\dot{p}^i)_{\rm rad}$ есть приращение 4-импульса излучения за единицу собственного времени, что подтверждается интегрированием уравнения (\ref{eq:4b}) по $d^4r^i$: по всей области, занятой излучением, и по времени от минус бесконечности до текущего момента времени $t$: 
\begin{equation}
\int{ T^{i0}_{\rm rad}(t)dV}=\int^{\tau(t)}(\dot{p}^i)_{\rm rad}d\tau.
\end{equation}
В то же время на больших расстояниях именно $ T^{ij}_{\rm rad }$ определяет поток и угловое распределенин излучения (см \cite{bib:G}).  Выражение для $(\dot{p})_{\rm rad}$, разумеется, известно и дается формулами для электродипольного излучения \cite{bib:L}.  

\textbf{Для учета воздействия излучения на движение электрона}  Дирак предложил следующего типа расширение Гамильтоновой системы ~\cite{bib:D}: 
\begin{equation}
\dot{p}^i=\frac{e}{c}F^{ij}\dot{x}_{j}- (\dot{p})^i_{\rm rad},\qquad
\dot{x}^i=m^{-1}{p}^i+(\dot{x})^i_{\rm rad}, \label{eq:5}
\end{equation}
выражения для добавленных членов пока несущественны.  

Такая интерпретация теории Дирака не бесспорна.  Более того, разбиение традиционной 4-силы радиационного трения на два слагаемых, $m\ d(\dot{x})^i_{\rm rad}/d\tau-(\dot{p})^i_{\rm rad}$,  (в используемых здесь обозначениях) и включение первого из них, являющегося полной производной по времени,  в определение  4-импульса, а не в уравнение импульсов, критиковалось, например, в \cite{bib:G}.  Однако, во-первых, Дирак ясно подчеркивал, что происхождение тех членов, что включены в энергию-импульс электрона в (\ref{eq:5}), связано именно с внутренней энергией электрона. Во-вторых, как показано выше, отделение этих членов от излучательных эффектов возможно и целесообразно, а вот корректное отделение их от энергии собственного поля электронов - вряд ли возможно.   

Выражая $(\dot{x})^i_{\rm rad}$ либо через $\dot{x}$, либо через $p^i$, можно рассматривать второе из уравнений (\ref{eq:5}) как операторную перенормировку либо массы: $p^i=\hat{m}[\dot{x}^j]$, где $ \hat{m}[\dot{x}^j]=m(\dot{x}^i+(\dot{x})^i_{\rm rad}[\dot{x}])$, либо обратной массы:  $\dot{x^i}=\hat{m}^{-1}[p^j]$, где $\hat{m}^{-1}[p^j]=m^{-1}p^i-(\dot{x})^i_{\rm rad}[p^j]$. 

\textbf{По сравнению с (\ref{eq:1}), свойства уравнений (\ref{eq:5}) изменяются} следующим образом. В законе сохранения энергии-импульса, 
\begin{equation}\label{eq:5a}
\dot{p}^i+(\dot{p})^i_{\rm rad}+\frac{d}{d\tau}\int{{T}^{i0}dV}=0,
\end{equation}
новый член,  $(\dot{p})^i_{\rm rad}$, описывает энергию и импульс, уносимые излучением.  Соотношение (\ref{eq:2}) остается неизменным, причем вклад от тока перехода,  $eF^i_{\ j}(\dot{x})^j_{\rm rad}/c$, имеет смысл энергии-импульса, которые электрон отбирает от поля в процессе излучения. Как поясняющий пример, рассмотрим релятивистски сильную электромагнитную волну  с волновым вектором $k^i_0$, в качестве внешнего электромагнтного поля. При излучении электроном в поле волны одного высокоэнергичного фотона c волновым вектором, $k^i_1$, закон сохранения энергии-импульса дает: $p^i_f=p^i_i+n\hbar k^i_0-\hbar k^i_1$,  нижний индекс, $i,f$, обозначает начальное и конечное состояние электрона. В сильном поле число поглощенных квантов, $n$, может быть очень велико, и их отбор от классического поля, уже хотя бы по этой причине, должен описываться в соответствии с уравнением (\ref{eq:2}) как действие тензора поля на некоторый ток. Именно этот ток и введен как $e(\dot{x})^i_{\rm rad}$. Импульс же излученного фотона дает вклад в $(\dot{p})^i_{\rm rad}$. Переход к пределу классической электродинамики приводит к уравнению типа (\ref{eq:5}) - а точнее, дает альтернативный вывод системы (\ref{eq:11}).

Закон сохранения обобщенного импульса видоизменяется, если импульс излучения имеет ненулевую проекцию на направление циклической координаты: $({\bf n}\cdot\dot{\bf P})=-{\bf n}\cdot(\dot{\bf  p})_{\rm rad}=0$. В отсутствие излучения и внешнего поля механический и обобщенный импульс сохраняются: в частности, равны нулю, если они изначально равны нулю (самовоздействие отсутствует). 

\textbf{Трудности однако возникают с выполнением уравнений (\ref{eq:3},\ref{eq:4}).} Отмеченная Дираком неоднозначность выбора выражения для импульса (иными словами, для перенормировки оператора массы), сопряжена с выбором - каким из этих уравнений следует пожертвовать, поскольку для системы (\ref{eq:5})  их одновременное выполнение невозможно.  В частности, обычно обеспечивается выполнение условия (\ref{eq:3}), что возможно, если $(\dot{x})^i_{\rm rad}\dot{x}_i=0$ и $m(\dot{x})^i_{\rm rad}\ddot{x}_i=-(\dot{p})^i_{\rm rad}\dot{x}_i $. Если излучательные потери выражены через ускорение согласно формуле для дипольного излучения, отсюда следует:
\begin{equation}
(\dot{p})^i_{\rm rad}=-\frac{m \tau_0\ddot{x}^2}{c^2}\dot{x}^i,
\qquad (\dot{x})^i_{\rm rad}=\tau_0\ddot{x}^i,\label{eq:6}
\end{equation}
\begin{equation}
\hat{m}[\dot{x}^i]=m(1-\tau_0\frac{d}{d\tau})\dot{x}^i.\label{eq:7} 
\end{equation}
Уравнения (\ref{eq:5},\ref{eq:6}) эквивалентны уравнению ЛАД:
\begin{equation}
m\ddot{x}^i=\frac{e}c F^{ij}\dot{x}_{j}+m\tau_0\dddot{x}^i +\frac{m \tau_0\ddot{x}^2}{c^2}\dot{x}^i.
\end{equation}
Отличие такой формы записи (см \cite{bib:D},\cite{bib:J}) от принятой в \cite{bib:L} не может вызвать недоразумение.

Общеизвестные трудности уравнения ЛАД ~\cite{bib:L}-\cite{bib:G} связаны с сингулярностью массового оператора: для самоускоряющегося решения $\dot{\bf x}\propto\exp(\tau/\tau_0)$, при том что скорость неограниченно растет,  импульс электрона равен нулю (см. (\ref{eq:7})).  Проблемой является и нарушение тождества (\ref{eq:4}), которое принимает вид: $p^2=m^2(c^2+\tau_0^2\ddot{x}^2)\le m^2c^2$ - напомним, что $\ddot{x}^2\le 0$.  Поскольку $p^2={\cal E}^2/c^2-{\bf p}^2$ , где ${\cal E}$ - энергия электрона, не только нарушается известное соотношение между энергией и импульсом: ${\cal E}^2/c^2\ne{\bf p}^2+m^2c^2$ , но и энергия покоя ускоренного электрона оказывается меньшей, чем $mc^2$.  Более того теория допускает превращение этого дефекта энергии в излучение: при анализе баланса (сохраняющейся!) энергии в самоускоряющемся решении легко видеть, что энергия излучения неограниченно черпается из энергии покоя электрона. 

При всех проблемах с энергией в уравнении ЛАД, отметим, что в известной приближенной модели для радиационной силы \cite{bib:L} вообще не просматривается возможность ввести энергию-импульс излучающей частицы, поскольку эта модель непредставима в виде (\ref{eq:5}).  Приближенная модель, представимая в виде (\ref{eq:5}), очень давно была предложена в \cite{bib:E}: 
\begin{equation}
(\dot{p})^i_{\rm rad}=-\frac{e \tau_0\ddot{x}_lF^{lj}\dot{x}_j}{c^3}\dot{x}^i,
\qquad (\dot{x})^i_{\rm rad}=\tau_0\frac{eF^{ij}\dot{x}_j}{mc},\label{eq:7a}
\end{equation}
\begin{equation}
\hat{m}[\dot{x}^i]=m^{ij}\dot{x}_j=m\left(g^{ij}-\tau_0\frac{eF^{ij}}{mc}\right)\dot{x}_j.\label{eq:7b} 
\end{equation}
Модель эта впоследствии была признана неудачной (см \cite{bib:K}). Однако дальнейшее преобразование уравнений (\ref{eq:5}, \ref{eq:7a}), сводящее их  к уравнению (76.3) из \cite{bib:L}, приводит к неточному сохранению энергии-импульса. Действительно,  преобразованный член $m\frac{d}{d\tau}(\dot{x})_{\rm rad}$ после выполнения дифференцирования по времени и приближенного выражения ускорения через силу Лоренца (как в \cite{bib:L})  уже не является полной производной по $\tau$, так что (\ref{eq:5a}) уже не выполняется ни в дифференциальной форме, ни даже в интегральной.

\textbf{Теперь потребуем выполнения (\ref{eq:4})  вместо (\ref{eq:3}),} что возможно в случае:
\begin{equation}
\frac{e}cp^iF_{ij}(\dot{x})^j_{\rm rad}=p_i(\dot{p})^i_{\rm rad}.\label{eq:8}
\end{equation}
Предполагая обычную связь между энергией и импульсом излучения ~\cite{bib:L},  получим согласованное с (\ref{eq:8}) выражение для тока перехода:
\begin{equation}
(\dot{p})^i_{\rm rad}=\frac{I}{mc^2}p^i, \qquad  (\dot{x})^i_{\rm rad}=\frac{\tau_0}{m}\frac{I}{I_E}f_L^i, \label{eq:9}
\end{equation}
 где $I$ - интенсивность излучения (потеря энергии за единицу времени), а $I_E=-\tau_0f^2_L/m$ - интенсивность электродипольного излучения, выраженная, как и сила Лоренца, $f_L^i=eF^{ij}p_j/(mc)$,  через импульс электрона и действующие на него поля. Соответственно перенормированный оператор (в данном случае - тензор) массы выглядит следующим образом:
\begin{equation}
(\hat{m}^{-1})^{ij}=\frac1mg^{ij}+\frac{\tau_0 e}{m^2c}\frac{I}{I_E}F^{ij},\label{eq:10}
\end{equation}
Если положить $I=I_E$ (что необязательно, $I$ может быть иной фукнцией или вообще не функцией, а случайной величиной, отражая вероятностный характер излучения кванта), приходим к системе уравнений движения для излучающего электрона:
\begin{equation}
\dot{p}^i=\frac{e}{c}F^{ij}\dot{x}_{j} +\frac{\tau_0f^2_L}{m^2c^2}p^i,\qquad
\dot{x}^i=m^{-1}{p}^i+\frac{\tau_0}{m}f_L^i. \label{eq:11}
\end{equation}
\textbf{Обсудим свойства полученных уравнений.} Прежде всего их отличие от уравнения ЛАД очень мало, что видно из сравнения (\ref{eq:5}-\ref{eq:6}) с (\ref{eq:11}), с учетом малости $\tau_0$. Тем не менее отсутствуют самоускоряющиеся решения и энергетические парадоксы.  Радиационные ``поправки'' к Гамильтоновой системе выражены через параметры ``невозмущенной'' системы (импульс и поле в точке, где находится частица), что допускает сопоставление с построенной по теории возмущений квантовой электродинамикой.  Полезна и проста трех-векторная формулировка (\ref{eq:11}):
\begin{align}
\frac{d{\bf p}}{dt}&={\bf f}_L+\frac{e}c[\delta{\bf u}\times{\bf B}]-\frac{{\cal E}^2(\delta{\bf u}\cdot{\bf f}_L)}{m^2c^6}{\bf u},\label{eq:12}\\
\frac{d{\bf x}}{dt}&={\bf u}+\delta{\bf u},\qquad  \delta{\bf u}=\frac{\tau_0}m\frac{{\bf f}_L-{\bf u}({\bf u}\cdot{\bf f}_L)/c^2}{1+\tau_0({\bf u}\cdot{\bf f}_L)/(mc^2)},\label{eq:13}
\end{align}
где $t$ - время в произвольной системе отсчета, ${\cal E}/c=(m^2c^2+{\bf p}^2)^{1/2}$, ${\bf u}=c^2{\bf p}/{\cal E}$, ${\bf f}_L=e({\bf E}+[{\bf u}\times{\bf B}]/c)$, - сила Лоренца, ${\bf E}$ и ${\bf B}$- электрическое и магнитное поля.

Численное решение уравнений (\ref{eq:12},\ref{eq:13}) в рамках метода частиц в ячейке не представляет труда и немногим сложнее чем для Гамильтоновых уравнений. При расчетах в рамках этого метода ток электронов со скоростью $\delta {\bf u}$, должен учитываться в уравнениях для электромагнитного поля, так как именно этот ток отвечает за отбор энергии от поля электроном в процессе излучения. Кроме того не представляет труда и подсчет полной энергии излучения, при этом может использоваться дополнительное к  (\ref{eq:12},\ref{eq:13})  уравнение энергии:
\begin{equation}
\frac{d{\cal E}}{dt}=e{\bf E}\cdot\left({\bf u}+\delta{\bf u}\right)-\frac{{\cal E}^2(\delta{\bf u}\cdot{\bf f}_L)}{m^2c^4}.\label{eq:14}
\end{equation}
Для излучаемой за единицу времени энергии, которая дается последним членом в (\ref{eq:14}), нетрудно также рассчитать угловое и спектральное распределение. Пример расчета по такой схеме дан в \cite{bib:N},\cite{bib:S}.
 
\textbf{Неизбежен вопрос о нарушении тождества (\ref{eq:3}).} Игнорируя благоприятные для теории моменты (при ограничении на величину поля, необходимом для применимости классической теории,  нарушение соотношения (\ref{eq:3}) мало: 
$$\frac{\hbar^2| f^2_L|}{m^2c^2}\ll m^2c^4,\qquad c^2(1-1/137^2)<\dot{x}^2\le c^2,$$
что во всяком случае исключает сверхсветовые парадоксы), рассмотрим главную логическую трудность: при нулевом пространственном импульсе электрона, ${\bf p}=0$, скорость излучающего электрона отлична от нуля и равна, как нетрудно видеть, $\dot{\bf x}_{\rm rad}=\tau_0e{\bf E}/m$ (именно поэтому $\dot{x}^2=c^2-\dot{\bf x}^2_{\rm rad}\le c^2$). Опять, игнорируем тот благоприятный факт, что за любой мыслимый интервал времени электрон наберет многократно большую скорость за счет ускорения полем.

Вынужденно прибегая к квантовым понятиям и используя приведенный выше пример электрона в поле одномерной релятивистски сильной электромагнитной волны, ответим последовательно на три вопроса. Что означает равный нулю (в течение некоторого малого интервала времени) импульс электрона? Он означает что если в течение этого времени излучение (вероятностный процесс) отсутствовало, то как импульс, так и скорость частицы в данной системе отсчета равны нулю или, в любой другой системе отсчета, электрон движется по траектории Гамильтоновой системы (\ref{eq:1}). Почему в процессе излучения скорость электрона становится отличной от нуля? Потому что, как объяснялось выше, для излучения электрон должен отобрать энергию от поля. Наконец, почему не допустить, что в процессе излучения импульс изменяется в соответствии со скоростью? Потому что это противоречило бы одному из законов сохранения: чтобы отобрать энергию от поля волны, скорость должна быть параллельна ${\bf E}$, но это направление циклической координаты, и изменение проекции импульса на это направление контролируется законом сохранения обобщенного импульса.  

Не видно иного выхода, как допустить нарушение соотношения (\ref{eq:3}), в качестве платы за непротиворечиво введенные энергию-импульс. Разумеется, непривычно представление об электроне, движущемся не строго в направлении импульса, или о ненулевой пространственной скорости при нулевом простраственном импульсе. Однако и то, и другое - всего лишь инверсии утверждения о том, что из-за эффекта запаздывания вклад собственного поля электрона в его импульс не может определяться мгновенным значением скорости заряда. 

\textbf{Пример аналитического решения найденных уравнений} легко получить для одномерного движения\footnote{Я признателен неизвестному рецензенту, указавшему на эту возможность} в однородном (для простоты) электрическом поле.  Координатную ось $x$ направляем вдоль $e{\bf E}$. Пусть в момент времени $t=\tau=0$ электрон имеет нулевой импульс и находится в точке $x=0$. Решение уравнений (\ref{eq:11}) имеет вид (сравните с \cite{bib:L}, \S 20,  \cite{bib:J}, задача 17.5 и \cite{bib:G}, Гл.3):
\begin{equation}
\frac{\cal E}{mc^2}=\cosh(\omega_E\tau),\qquad \frac{p_x}{mc}=\sinh(\omega_E\tau),\qquad \omega_E=\frac{eE}{mc},\label{eq:15}
\end{equation} 
причем в пренебрежении излучением эволюция энергии-импульса имела бы точно такой же характер.  Однако движение излучающей частицы имеет существенное отличие:
\begin{equation}
\frac{x}c=\frac{\cosh(\omega_E\tau)-1}{\omega_E } +\tau_0\sinh(\omega_E\tau),
\qquad t=\frac{\sinh(\omega_E\tau)}{\omega_E } +\tau_0\left[\cosh(\omega_E\tau)-1\right].\label{eq:16}
\end{equation}
По сравнению с движением которое имело бы место в отсутствне излучения, то есть в пренебрежении $\propto\tau_0$ членами в (\ref{eq:16}),  излучающий электрон проскальзывает вперед в пространстве и во времени ($\propto\tau_0$ члены оба положительны). При таком проскальзывании электрическое поле совершает дополнительную работу, которая компенсирует излученную энергию и импульс:
\begin{equation}
{\cal E}_{\rm rad}=\tau_0\omega_E\sinh(\omega_E\tau),\qquad
(p_x)_{\rm rad}=\tau_0\omega_E\left[\cosh(\omega_E\tau)-1\right].
\end{equation}
Выбирая различную калибровку (чисто скалярный потенциал или чисто векторный), нетрудно проследить сохранение не только энергии, но и обобщенного импульса. Другое аналитическое решение (для электрона в поле одномерной электромагнитной волны) дано в \cite{bib:S}.

Работа автора по физике высоких плотностей энергии поддержана грантом DE-FC52-08NA28616.

\end{document}